\RequirePackage{fix-cm}
\documentclass[smallextended]{svjour3}       

\smartqed  
\usepackage{graphicx}

\usepackage{amssymb}
\usepackage{amsmath}
\usepackage{mathtools}
\usepackage{bigdelim}
\usepackage{algpseudocode}

\begin{document}

\title{Globally Optimal Cooperation in Dense Cognitive Radio Networks
}
\subtitle{}


\author{Ahmed M. Alaa         \and
        Omar A. Nasr 
}


\institute{Ahmed M. Alaa \at
              Electronics and Electrical Communications Engineering Dept., Cairo University,\\
              Gizah 12316, Egypt \\
              Tel.: +20-0100-2968798\\
              \email{aalaa@eece.cu.edu.eg}           
           \and
           Omar A. Nasr \at
					 Electronics and Electrical Communications Engineering Dept., Cairo University,\\
           Gizah 12316, Egypt \\
           \email{omaranasr@ieee.org}
}

\date{Received: date / Accepted: date}

\maketitle

\begin{abstract}
The problem of calculating the local and global decision thresholds in hard decisions based cooperative spectrum sensing is well known for its mathematical intractability. Previous work relied on simple suboptimal counting rules for decision fusion in order to avoid the exhaustive numerical search required for obtaining the optimal thresholds. However, these simple rules are not globally optimal as they do not maximize the overall global detection probability by jointly selecting local and global thresholds. Instead, they maximize the detection probability for a specific global threshold. In this paper, a globally optimal decision fusion rule for Primary User signal detection based on the Neyman-Pearson (NP) criterion is derived. The algorithm is based on a novel representation for the global performance metrics in terms of the regularized incomplete beta function. Based on this mathematical representation, it is shown that the globally optimal NP hard decision fusion test can be put in the form of a conventional one dimensional convex optimization problem. A binary search for the global threshold can be applied yielding a complexity of $\mathcal{O}(\log_{2}(N))$, where $N$ represents the number of cooperating users. The logarithmic complexity is appreciated because we are concerned with dense networks, and thus $N$ is expected to be large. The proposed optimal scheme outperforms conventional counting rules, such as the OR, AND, and MAJORITY rules. It is shown via simulations that, although the optimal rule tends to the simple OR rule when the number of cooperating secondary users is small, it offers significant SNR gain in dense cognitive radio networks with large number of cooperating users.  
\end{abstract}

\section{Introduction}
\label{intro}
Cognitive radio (CR) is a promising technology offering enhanced spectrum efficiency via dynamic spectrum access [1], [2]. In a CR network, unlicensed Secondary Users (SU) can opportunistically occupy the unused spectrum allocated to a licensed primary user (PU). This is achieved by means of PU signal detection. Detection of PU signal entails sensing the spectrum occupied by the licensed user in a continuous manner. Based on the sensing data, the SU is required to decide whether or not a PU exists. 
A common problem encountered in CR systems is the \textit{hidden terminal problem} [3], where shadowing and multipath fading affect the strength of the PU signal causing it to be undetectable. Hence, spatial diversity is applied by utilizing multiple decisions from several SU terminals using a decision fusion rule. The fusion rule is applied by a central terminal known as the \textit{fusion center}. 
Two basic approaches for decision combining are discussed in literature: \textit{soft decision} (SD) and \textit{hard decision} (HD) combining. The former relies on adding the sensed energies, while the latter combines one-bit local decisions to make a final decision [4]. 

In this work, we tackle the problem of optimizing the HD combining scheme based on Neyman-pearson (NP) criterion. While the optimal NP test has been formulated for the SD combining case [4], it is much more challenging to apply an optimal NP test for the HD combining scheme. The reason for this is that every SU employs a local detection threshold, while the fusion center applies a global threshold to make a final decision on the existence of a PU. Thus, unlike the simple one-dimensional problem in SD combining, two degrees of freedom are considered in the HD combining optimization problem. In his pioneering work, Tsitsiklist [5] showed that the problem is mathematically intractable and an exhaustive search would be used to obtain local detection thresholds. In a recent comprehensive survey, Quan \textit{et al} [6] pointed out that computing the optimal decision thresholds
under the NP criterion is mathematically intractable. Various suboptimal solutions were presented in literature. In [7], the problem was solved by simply fixing local thresholds and obtaining the optimum global threshold or vice versa. Recently, the problem was revisited in [8], were large deviation analysis was used to determine a local decision rule to optimize the asymptotic global
performance. However, the intractability of the exact NP optimization problem was again emphasized. In literature, the adopted HD combining rules are never globally optimal. Researchers usually employ simple suboptimal AND, OR or MAJORITY counting rules for global detection [9][10]. Others try to calculate the optimim local and global thresholds but mainly using exhaustive numerical methods [11][12]. In [4], the performance of the SD combining scheme with NP test was compared with an OR-rule based HD combining scheme, which is not necessarily optimal. The problem of HD and SD performance comparison was thoroughly studied in [12]. However, the authors used suboptimal counting rules and stated that the threshold calculations are not trivial as complex optimization schemes are needed to solve them.

Although simple fusion rules, such as the OR rule, is usually found to be optimal for cognitive radio networks with small number of cooperating SUs, it was never verified in literature that the same applies for dense networks with large number of SUs. In this work, we propose a globally optimal decision fusion rule for HD combining based on Neyman-pearson criterion. It is shown that the NP optimal thresholds can be obtained by solving a simple one-dimensional convex optimization problem. Besides, we obtain a closed form expression for the local detection threshold as a function of the optimal global threshold. A simple and efficient algorithm for optimizing global and local thresholds is proposed. Although the algorithm is general and can be applied for any number of SUs, it is shown that it offers significant performance gain compared to the OR rule in networks with large number of cooperating users. 

The rest of the paper is organized as follows. In section 2, we present the system model. Next, we propose the globally optimal HD combining scheme in section 3. Simulation results are discussed in section 4. Finally, we draw our conclusion in section 5.

\begin{flushright}
\section{System Model}

\begin{flushright}

\end{flushright}

We investigate cooperative spectrum sensing in a CR network with $N$ cognitive users and a single common receiver (Fusion Center). We assume that the SU observes $M$ samples for spectrum sensing. Energy detection is adopted as a spectrum sensing technique. It is assumed that the instantaneous SNR at the $j^{t h}$ node is $\gamma_{j}$ and the primary signalӳ $i^{t h}$ sample at the $j^{t h}$ CR is $S_{j i}$, and considered constant with unity power for the entire sensing period. The additive white noise is $n_{j i}$ $\sim$ $\mathcal{N}(0, 1)$. Thus, the $i^{t h}$ sample received at the $j^{t h}$ CR is a binary hypothesis give by:  
\begin{equation}
\label{eqn_example}
   r_{j i} = \left\{
     \begin{array}{lr}
       n_{j i}, & \ \mathcal{H}_{o} \\
       \sqrt{\gamma_{j}} \hspace{1 mm} S_{j i} + n_{j i}, & \ \mathcal{H}_{1} 
     \end{array}
   \right.
\end{equation} 

The conditional distributions on null and alternative hypotheses are:
\begin{equation}
\label{eqn_example}
   r_{j i} \sim \left\{
     \begin{array}{lr}
       \mathcal{N}(0, 1), & \ \mathcal{H}_{o} \\
       \mathcal{N}(\sqrt{\gamma_{j}}, 1), & \ \mathcal{H}_{1} 
     \end{array}
   \right.
\end{equation} 

where $\mathcal{H}_{o}$ denotes the absence of the PU, while $\mathcal{H}_{1}$ denotes the existence of the PU. After applying such signal to an \textit{energy detector} and obtaining binary decisions on PU existence, the local false alarm and detection probabilities at the $j^{t h}$ CR are [2]: 
\[P_{F}(M,\lambda) = P(Y_{j} > \lambda | \mathcal{H}_{o}) = \frac{\Gamma (\frac{M}{2},\frac{\lambda}{2})}{\Gamma(\frac{M}{2})},\]
and
\begin{equation}
\label{eqn_example} 
P_{D}(M,\lambda,\gamma_{j}) = P(Y_{j} > \lambda | \mathcal{H}_{1}) = Q_{M/2}(\sqrt{2\gamma_{j}}, \sqrt{\lambda})
\end{equation}
where $\lambda$ is the local threshold, $\Gamma(.,.)$ is the incomplete gamma function, $\Gamma(.)$ is the gamma function, and $Q_{u}(.)$ is the generalized Marcum Q-function. We assume Rayleigh fading with an average SNR of $\overline{\gamma}$. The average SNR is assumed to be the same for all CR users. The instantaneous SNR is assumed to be constant over the $M$ observable samples. Different observations perceive different SNR values. The SNR varies according to the exponential pdf:
\begin{equation}
\label{eqn_example} 
f_{\gamma}(\gamma) = \frac{1}{\overline{\gamma}} e^{-\frac{\gamma}{\overline{\gamma}}}, \gamma \geq 0.
\end{equation}
The reporting channel between the SUs and the fusion center is assumed to be free of errors. 

\end{flushright}

\section{Globally Optimal Hard Decision Fusion}
In this section, we propose a globally optimal algorithm for HD combining based on the Neyman-Pearson criterion. The ultimate goal of a Neyman-pearson test is to maximize the detection probability for a given false alarm probability. The overall performance of the HD scheme is determined by the global detection and false alarm probabilities, which are functions of the local detection and false alarm probabilities given in equation (3). As the fusion center employs an \textit{n-out-of-N rule} fusion rule, we let $l$ be the test statistic denoting the number of votes for the existence of PU from the $N$ SU votes. Hence, the conditional pdfs follow the \textit{binomial distribution} as [3]:
\[P(l | \mathcal{H}_{o}) = \binom{N}{l} \hspace{1.5 mm} P_{F}^{l} \hspace{1.5 mm} (1-P_{F})^{N-l}\]
and
\begin{equation}
\label{eqn_example} 
P(l | \mathcal{H}_{1}) = \binom{N}{l} \hspace{1.5 mm} \overline{P}_{D}^{l} \hspace{1.5 mm} (1-\overline{P_{D}})^{N-l},
\end{equation}

where $\overline{P}_{D}$ is the local detection probability averaged over the fading channel pdf as follows:
\begin{equation}
\label{eqn_example} 
\overline{P}_{D} = \int_0^{\infty} Q_{M/2}(\sqrt{2\gamma}, \sqrt{\lambda}) \,\, \frac{1}{\overline{\gamma}} e^{-\frac{\gamma}{\overline{\gamma}}}  \,d\gamma.
\end{equation}
and the global false alarm and detection probabilities $Q_{f}$ and $Q_{d}$ are [3][12]:
\[Q_{f}(n,\lambda)= \sum_{l=n}^{N} \binom{N}{l} \hspace{1.5 mm} P_{F}^{l}(\lambda) \hspace{1.5 mm} (1-P_{F}(\lambda))^{N-l},\]

\begin{equation}
\label{eqn_example} 
Q_{d}(n,\lambda) = \sum_{l=n}^{N} \binom{N}{l} \hspace{1.5 mm} \overline{P}_{D}^{l}(\lambda) \hspace{1.5 mm} (1-\overline{P}_{D}(\lambda))^{N-l}.
\end{equation}
The global Neyman-pearson threshold for the discrete observable random variable $l$ is denoted by $n$. We search for the pair of thresholds $(n, \lambda)_{opt}$ that maximizes the global detection probability $Q_{d}$ for $Q_{f}$ = $\alpha$. Unlike the conventional Neyman-Pearson detection schemes, we have two degrees of freedom dictated by the local and global thresholds. 

The cumulative density function (CDF) of the binomial distribution can be written in the form of the \textit{regularized incomplete beta function} defined as [13, eq. 6.6.2]: 
\[\mathcal{I}(x;a,b) = \frac{\beta(x;a,b)}{\beta(a,b)},\]
where $\beta(x;a,b) = \int_0^{x} t^{a-1} (1-t)^{b-1} dt$ is the upper incomplete beta function and $\beta(a,b) = \int_0^{1} t^{a-1} (1-t)^{b-1} dt$ is the beta function. The CDF of a binomial random variable $x \sim B(N, p)$ is $F(x \leq X) = \mathcal{I}(1-p; N-X, X+1)$ [13, eq. 6.6.4]. Thus, the cumulative density of the discrete variable $l$ under $\mathcal{H}_{o}$ hypothesis is given by:   
\begin{equation}
\label{eqn_example} 
P(L \leq n | \mathcal{H}_{o}) = \mathcal{I}(1-P_{F};N-n, n+1),
\end{equation}
and the global false alarm probability is given by:
\begin{eqnarray}
Q_{f} & = & 1-P(L<n | \mathcal{H}_{o}) = 1 - P(L \leq n-1 | \mathcal{H}_{o})  \notag \\ 
             & = & 1-\mathcal{I}(1-P_{F};N-n+1, n). 
\end{eqnarray}
One of the properties of the regularized incomplete beta function is the \textit{symmetry property} [13, eq. 6.6.3]:
\[1-\mathcal{I}(1-p;a,b) = \mathcal{I}(p,b,a).\]
Applying this property to equation (9):
\begin{equation}
\label{eqn_example} 
Q_{f} = \mathcal{I}(P_{F};n, N-n+1),
\end{equation}
and by using the \textit{inverse regularized beta function}, we can obtain the local false alarm probability by setting $Q_{f}$ = $\alpha$:
\begin{equation}
\label{eqn_example} 
P_{F} = \mathcal{I}^{-1}(\alpha;n, N-n+1).
\end{equation}
The regularized beta function and its inverse are implemented with low complexity algorithms in mathematical software tools like MATLAB and MATHEMATICA. The same algorithms can be implemented at the SU recievers. Similarly, the global detection probability is given by:
\begin{equation}
\label{eqn_example} 
Q_{d} = \mathcal{I}(\overline{P}_{D};n, N-n+1).
\end{equation}
Before presenting the proposed Neyman-Pearson algorithm, we construct some auxiliary mathematical tools. We define the functions $\zeta_{M}(x)$ and $\Phi_{M}(x,a,b)$ as: 
\[\zeta_{M}(x) = \frac{\Gamma (\frac{M}{2},\frac{x}{2})}{\Gamma(\frac{M}{2})}\]
and
\begin{equation}
\label{eqn_example} 
\Phi_{M}(x,a,b) = \mathcal{I}(\zeta_{M}(x);a, b-a+1).
\end{equation}
With the inverse function given by:
\begin{equation}
\label{eqn_example} 
\Phi_{M}^{-1}(y,a,b) = \zeta_{M}^{-1}(\mathcal{I}^{-1}(y;a, b-a+1)).
\end{equation}
Where $\zeta^{-1}_{M}(.)$ is the \textit{inverse incomplete gamma} function. We can rewrite the global false alarm probability and local threshold in terms of the $\Phi_{M}(x;a,b)$ function by combining equation (3) with equation (10):
\begin{equation} 
Q_{f} = \Phi_{M}(\lambda;n,N),
\end{equation}
\begin{equation} 
\lambda = \Phi_{M}^{-1}(\alpha;n,N).
\end{equation}
Note that equation (15) is a single equation in two unknowns $n$ and $\lambda$. Thus, there is an infinte number of $(n, \lambda)$ pairs that solve (15). We search for the pair that maximizes the expression in (12).
\def\minz{\hbox{\raise-2mm\hbox{$\textstyle min \atop \tiny{n} \in \{1,2,\cdots,N\}$}}}

\newcommand{\argmin}{\operatornamewithlimits{arg min}}

The global detection probability $Q_{d}(n)$ is a log-concave function of the global threshold $n$. Thus, the global and local threshold pair $(n,\lambda)_{opt}$ is obtained by solving the convex optimization problem: 
   \[n_{opt} = \underset{n \in \{1,\cdots,N\}}{\operatorname{arg\,min}} \, \biggl( -\ln \left(\hspace{1.25 mm} \mathcal{I}\left(\hspace{1 mm}\overline{P}_{D}(n);n, N-n+1 \right) \right) \biggr),\]
	and
	 \begin{equation}
			\lambda_{opt} = \Phi_{M}^{-1}(\alpha;n_{opt},N).
	 \end{equation}

Our objective is to prove that the global detection probability in equation (12) is a log-concave function of $n$. Hence, taking the negative of its natural logarithm leads to a straight forward convex optimization problem. Note that the regularized incomplete beta function can be written in terms of the \textit{gauss hypergeometric function} $_{2} F_{1} \left(.; .; .;. \right) $ as [14, eq. 8.392]:
\[Q_{d}(n) = \frac{{\overline{{P}}_{D}}^{n}}{n \hspace{1 mm} \beta(n, N-n+1)} \hspace{1 mm} _{2} F_{1} \left( n; n-N; n+1; \overline{P}_{D} \right). \]
Furthermore, the beta function can be obtained in terms of the gamma function as in [14, eq. 8.384.1] which yields:
\[Q_{d}(n) = \frac{{\overline{{P}}_{D}}^{n} \hspace{1 mm} \Gamma(N+1)}{n \hspace{1 mm} \Gamma(n) \Gamma(N-n+1)} \hspace{1 mm} _{2} F_{1} \left( n; n-N; n+1; \overline{P}_{D} \right). \]
By replacing the gauss hypergeometric function by the equivalent series representation [15, eq. (4)]:
\[Q_{d}(n) = \frac{{\overline{{P}}_{D}}^{n} \hspace{1 mm} \Gamma(N+1)}{n \hspace{1 mm} \Gamma(n) \Gamma(N-n+1)} \hspace{1 mm} \sum_{k=0}^{\infty} \frac{(n)_{k} (n-N)_{k}}{(n+1)_{k}} \times \frac{{\overline{{P}}_{D}}^{k}}{k!},\]
where $(a)_{k} = a(a + 1) \cdots (a + k - 1)$ is \textit{Pochhammer's symbol}, which can be represented by $(a)_{k} = \frac{\Gamma(a+k)}{\Gamma(a)}$ [15, eq. (1)]. By simplifying the above expression using the gamma function representation of the Pochhammer symbols, the function $Q_d(n)$ becomes: 
\[Q_{d}(n) = \sum_{k=0}^{\infty} \Xi(n,k), \]
where
\[\Xi(n, k) \propto \]
\begin{equation}
\underbrace{(n-N)_{k}}_{F_{1}(n,k)} \times \underbrace{\frac{1}{n \Gamma(n)}}_{F_{2}(n,k)} \times \underbrace{\frac{1}{(n+k) \Gamma(N-n)}}_{F_{3}(n,k)} \times \underbrace{{\overline{P}_{D}}^{n+k}}_{F_{4}(n,k)}.
\end{equation}
Thus, the global detection probability is composed of $\Xi(n, k)$ terms that are summed over $k$. Every $\Xi(n, k)$ term is proportional (within a positive scale) to the product of the terms $F_{1}(n,k)$, $F_{2}(n,k)$, $F_{3}(n,k)$ and $F_{4}(n,k)$ as depicted by equation (18). We start by studying the behavior of each $F(n,k)$ term individually.

\begin{itemize}
\item \textbf{log-concavity of $F_{1}(n,k)$} \\ In order to prove the log-concavity of Pochhammer's symbol $F_{1}(n,k) = (n-N)_{k}$, we take the natural logarithm of the gamma function representation of $F_{1}(n,k)$ as:
\[\ln(F_{1}(n,k)) = \ln(\Gamma(n-N+k)) - \ln(\Gamma(n-N)).\]
Applying the second derevative test, we get:
\[\frac{\partial^{2} \ln(F_{1}(n,k)) }{\partial n^{2}}= \psi^{\tiny{(1)}}(n-N+k) - \psi^{\tiny{(1)}}(n-N),\]
where $\psi^{\tiny{(1)}}(x)$ is the \textit{first order polygamma} function [13, eq. 6.4.1]. Based on the property $\psi^{(1)}(x+1) = \psi^{(1)}(x) - \frac{1}{x^{2}}$ [13, eq. 6.4.6], we conclude that $\psi^{(1)}(x+k) < \psi^{(1)}(x), \forall k > 0$. Thus, $\psi^{\tiny{(1)}}(n-N+k) - \psi^{\tiny{(1)}}(n-N)$ is always negative and the function $F_{1}(n,k)$ is log-concave.  

\item \textbf{log-concavity of $F_{2}(n,k)$ and $F_{3}(n,k)$} \\
The second derevative test for $F_{2}(n,k)$ is given by:
\[\frac{\partial^{2} \ln(F_{2}(n,k))}{\partial n^{2}} = \psi^{\tiny{(1)}}(n+1) - 2\psi^{\tiny{(1)}}(n),\]
which is always negative as $\psi^{(1)}(x+1) < \psi^{(1)}(x)$, $\forall x > 0$. Hence, the second derevative test shows the log-concavity of $F_{2}(n,k)$. A similar analysis can be applied to $F_{3}(n,k)$.

\begin{figure}
  \includegraphics[width=5in]{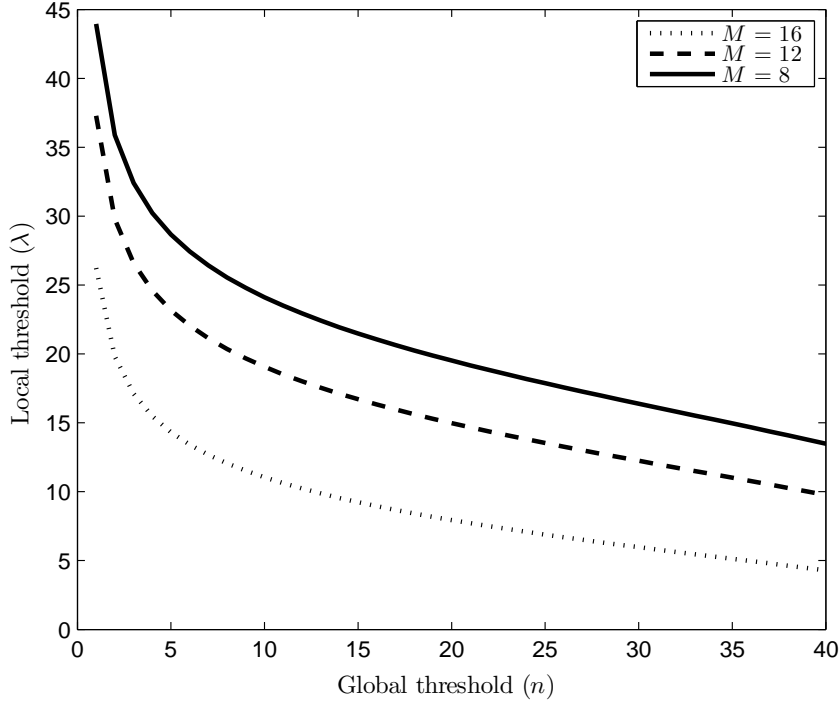}
\caption{The behavior of local threshold as a function of the global threshold.}
\label{fig:1}       
\end{figure}

\begin{figure}
  \includegraphics[width=5in]{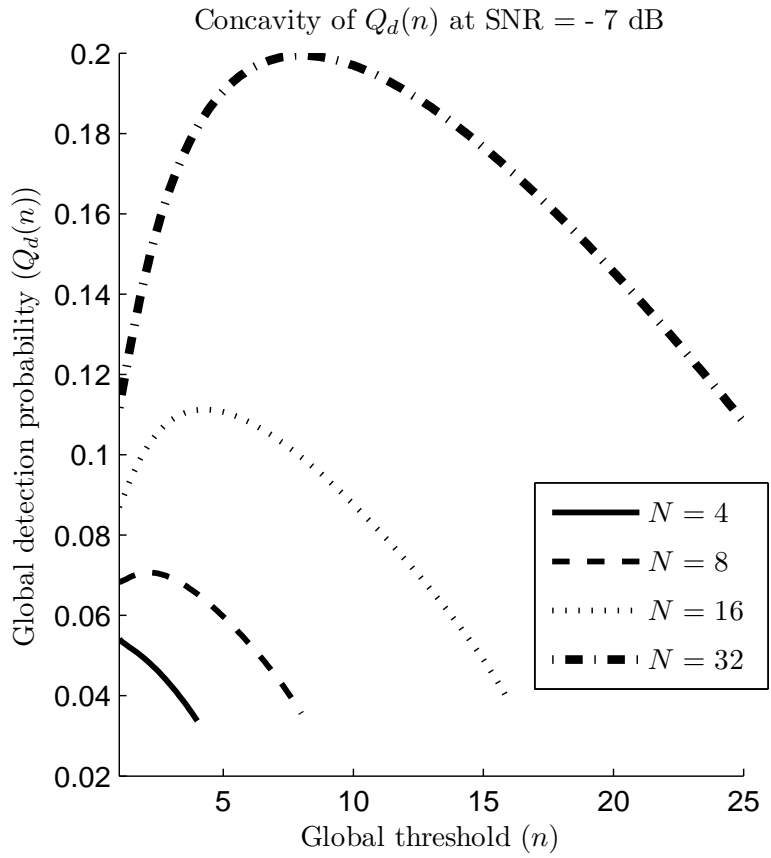}
\caption{The concavity of $Q_d(n)$ for various numbers of cooperating users.}
\label{fig:1}       
\end{figure}

\begin{figure}
  \includegraphics{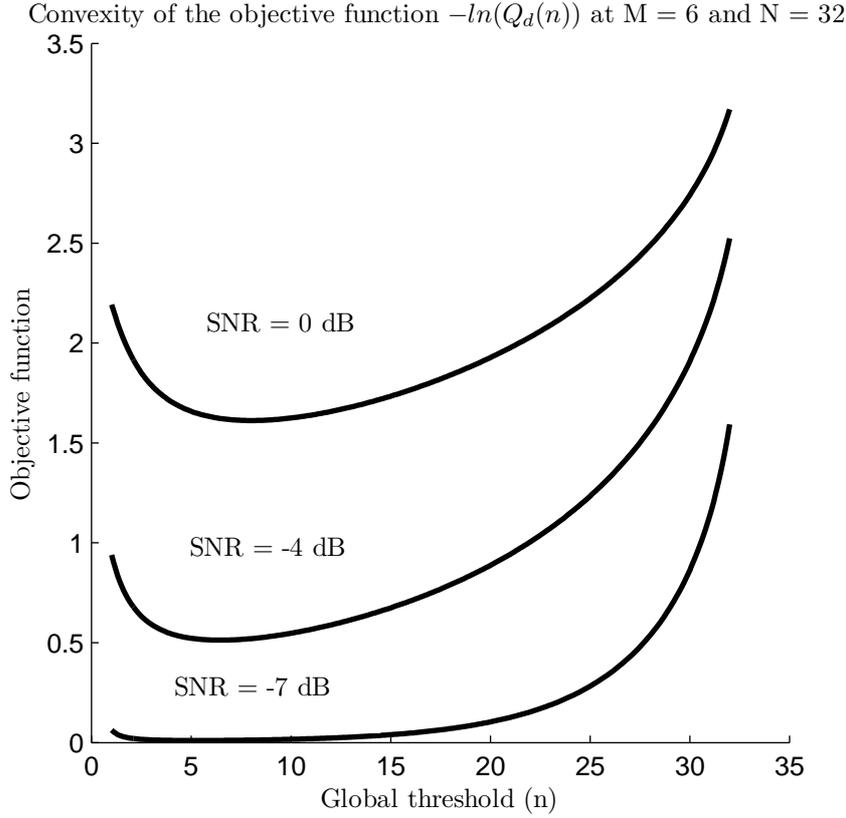}
\caption{Convexity of the objective function.}
\label{fig:1}       
\end{figure}

\item \textbf{log-concavity of $F_{4}(n,k)$} \\  
Note that $F_{4}(n,k)$ is given by 

\[F_{4}(n,k) = \int_0^{\infty} Q_{M/2}(\sqrt{2\gamma}, \sqrt{\lambda}) \,\, \frac{1}{\overline{\gamma}} e^{-\frac{\gamma}{\overline{\gamma}}}  \,d\gamma.\]

The log-concavity of the functions $b \to Q_{M/2}(a,b)$ and $b \to Q_{M/2}(a,\sqrt{b})$ were shown in [14]. Thus, $Q_{M/2}(\sqrt{2\gamma}, \sqrt{\lambda})$ is a log-concave function of $\lambda$. By discretization of the integral defining $F_{4}(n,k)$, we obtain

\[F_{4}(n,k) = \lim_{\bigtriangleup\gamma \to 0} \sum_{i=0}^{\infty} Q_{M/2}(\sqrt{2 i \bigtriangleup\gamma}, \sqrt{\lambda}) \,\, \frac{1}{\overline{\gamma}} e^{-\frac{i\bigtriangleup\gamma}{\overline{\gamma}}}  \,\bigtriangleup\gamma.\]
    
Because the terms $\frac{1}{\overline{\gamma}} e^{-\frac{i\bigtriangleup\gamma}{\overline{\gamma}}}  \,\bigtriangleup\gamma$ in the summation are all positive, and the terms $Q_{M/2}(\sqrt{2 i \bigtriangleup\gamma}, \sqrt{\lambda})$ are all log-concave in $\lambda$, thus $F_{4}(n,k)$ is the sum of positive scaled log-concave functions, which means that $F_{4}(n,k)$ is also a log-concave function. 	
		
\end{itemize}

Based on the above discussion, we conclude that the function $\Xi(n, k)$ is a product of log-concave functions. As the product and addition operations preserve log-concavity [17], $\Xi(n, k)$ and $Q_{d}(n)$ are both log-concave on all positive values of $n$. Because $Q_{d}(n)$ is a log-concave function of $n$, we can obtain the global threshold by minimization of the convex function $-\ln(Q_{d}(n))$.

To sum up, a cognitive radio user needs to perform a simple two step algorithm in order to obtain the optimal thresholds. Given $\overline{\gamma}$, $M$, $N$, and assuming that $N$ is odd, the SU applies the following two steps:
\\

\textbf{Step 1: Obtain the optimal global threshold $n_{opt}$ by applying convex minimization to the objective function $\biggl( -\ln \left(\hspace{1.25 mm} \mathcal{I}\left(\hspace{1 mm}\overline{P}_{D}(n);n, N-n+1 \right) \right) \biggr)$.} \\

This can be done using a binary search as follows:


\begin{algorithmic}[1]
\Procedure{Global threshold}{$N,M$}
\State $n_{opt}\gets 0$ 
\State $i\gets 1$
\State $j\gets \frac{N}{2}$
\State $k\gets 0$
\State $F(n) \gets \biggl( -\ln \left(\hspace{1.25 mm} \mathcal{I}\left(\hspace{1 mm}\overline{P}_{D}(n);n, N-n+1 \right) \right) \biggr)$
\While{$k\not=1$}

\If {$F(j) \leq F(j+1) \,\ and \,\ F(j) \leq F(j-1)$}
    \State $n_{opt} \gets j$
		\State $k \gets 1$
\Else
    \State $i \gets i+1$
	  \State $j \gets j+ sign(F(j-1)-F(j+1)) \frac{N}{2^{i}}$
\EndIf

\EndWhile\label{euclidendwhile}
\State \textbf{return} $n_{opt}$
\EndProcedure
\end{algorithmic}

\textbf{Step 2: Obtain the optimal local thresholds using the equation $\lambda_{opt} = \Phi_{M}^{-1}(\alpha;n_{opt},N)$.} \\

The optimization of the objective function is a done using a simple binary search approach. The feasibility of binary search is due to the convexity of the set of points representing the discrete objective function $-ln(Q_{d})$. Thus, the algorithm has a complexity of  $\mathcal{O}(\log_{2}(N))$, and it scales logarithmically with the number of cooperating users. Because we are mainly concerned with dense networks, the logarithmic complexity is appreciated. This would be appreciated by CR reciever designers as threshold optimization has to be done every time the listening or reporting channels change [12]. Figures 2 depicts the impact of the number of cooperating users and SNR on $Q_{d}(n)$ for a false alarm probability of 0.01. It is shown that as more users cooperate, the detection probability improves. It is found that an OR-rule would be optimal for the case of $N$ = 4 case. However, as $N$ increases, the maximum detection probability becomes interior to the range $(1,N)$. Figure 3 depicts the convexity of the objective function $-ln(Q_{d}(n))$ at $N$ = 32. It is shown that increasing SNR will normally lead to an enhanced detection performance.  

\section{Simulation results}

In this section, we aim at characterizing the performance of the proposed globally optimal algorithm. The optimal fusion rule employs the thresholds calculated via the optimization problem in (17). We first verify the accuracy of the analytic model adopted in our work. In figure 4, the simulated detection probability is plotted versus SNR and compared with the numerical results obtained from equation (12). It is shown that both results nearly coincide. In order to verify the optimality of the proposed algorithm, a comparison is done between the optimal rule and the conventional AND, OR and MAJORITY rules in figure 5. In all simulations, we set $Q_{F}$ = 0.01. It is shown that for N = 16, the optimal rule offers 1 dB SNR gain over the OR-rule and 1.5 dB gain over MAJORITY rule. The optimal scheme significantly outperforms the AND rule scheme. Moreover, the impact of the number of sensing samples $M$ (or equivalently, the sensing time) is demonstrated in figure 6. At an SNR of -2 dB and N = 16, we plot the global detection probability for $M$ = 6, 12, 18, and 24. It is shown that the maximum detection probability is significantly boosted from more than 0.6 at $M$ = 6 to more than 0.9 at $M$ = 18. This boost in detection probability comes on the expense of sensing delay. Figure 7 translates this detection probability boost into an SNR gain for the same number of cooperating users ($N$ = 16). It is found that increasing the number of sensing samples from 6 to 24 can offer up to a 4 dB SNR gain. It is worth mentioning that the proposed scheme offers significant gain only in networks with large number of cooperating users. As demonstrated by figure 8, when N = 8, the OR-rule and the optimal fusion rule have nearly equal performance. The attained SNR gain is only significant when the number of cooperating users increase to N = 16 and 32. The SNR gain attained in both cases are 1 dB and 2 dB respectively. Thus, the proposed scheme would be appreciated in dense cooperative networks.     

\begin{figure}[!h]
\centering
\includegraphics[width=5in]{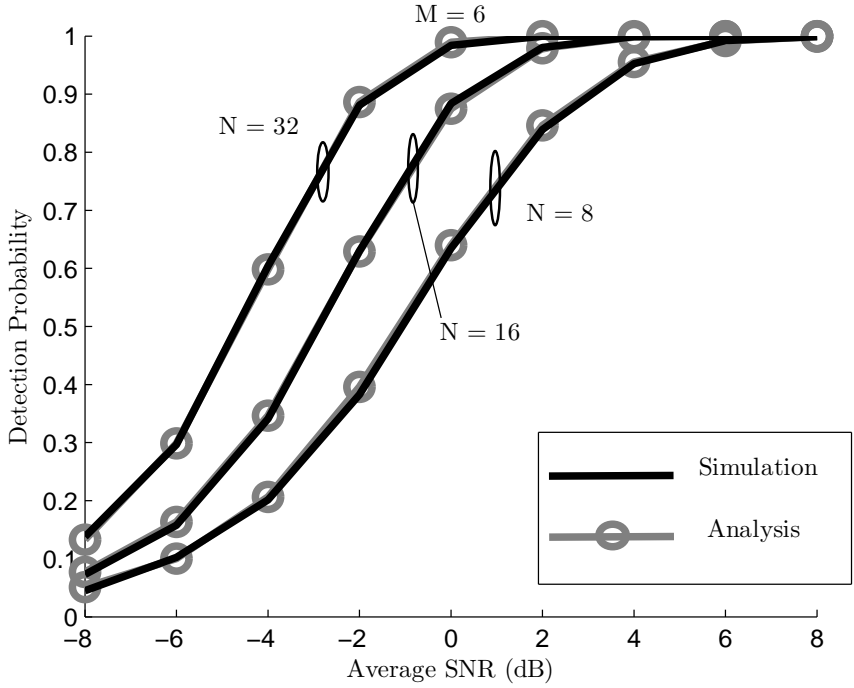}
\caption{Simulation results comapred with the proposed analysis.}
\label{fig_sim}
\end{figure}    

\begin{figure}[!h]
\centering
\includegraphics[width=5in]{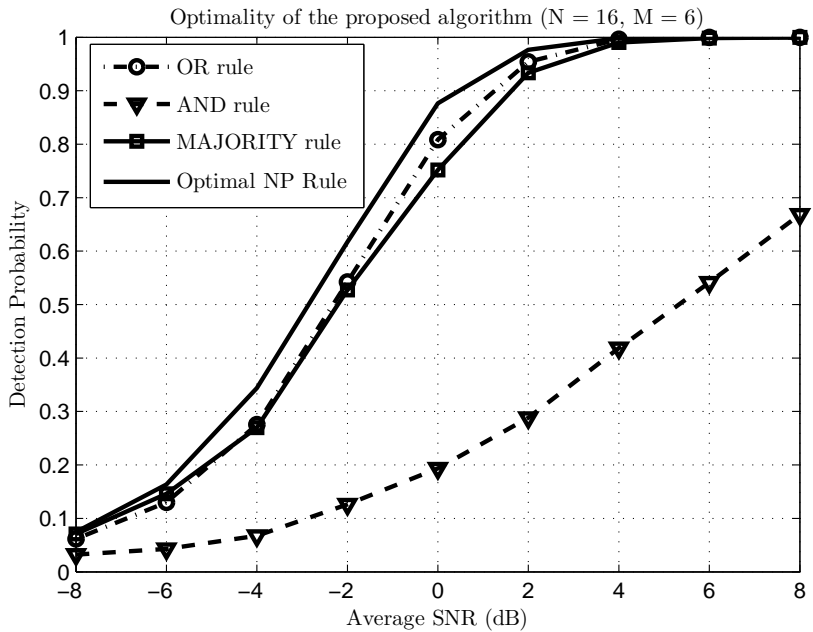}
\caption{Comparison between optimal rule and suboptimal counting rules.}
\label{fig_sim}
\end{figure}

\begin{figure}[!h]
\centering
\includegraphics[width=5in]{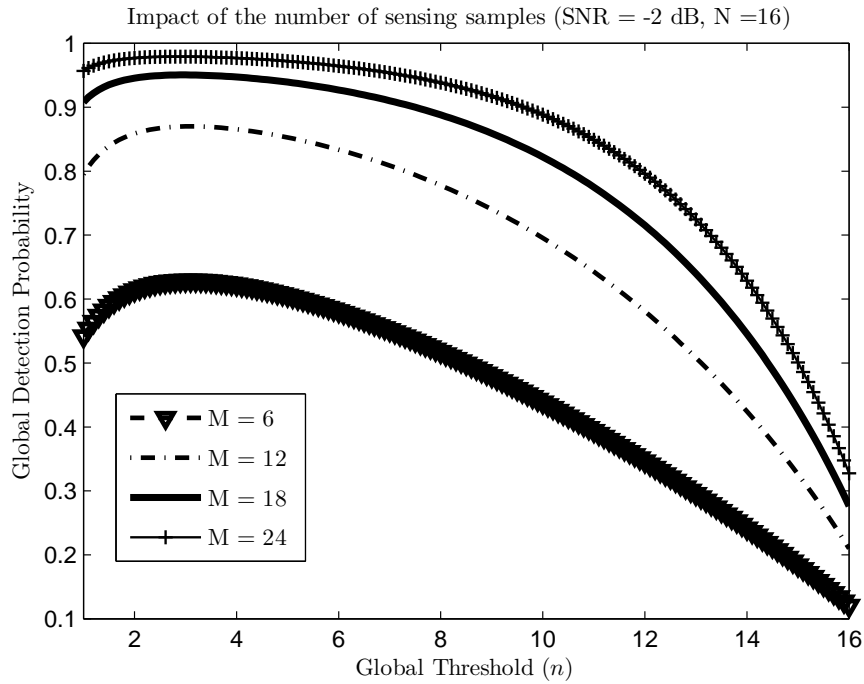}
\caption{Impact of the number of sensing samples on the global detection probability.}
\label{fig_sim}
\end{figure}

\begin{figure}[!h]
\centering
\includegraphics[width=5in]{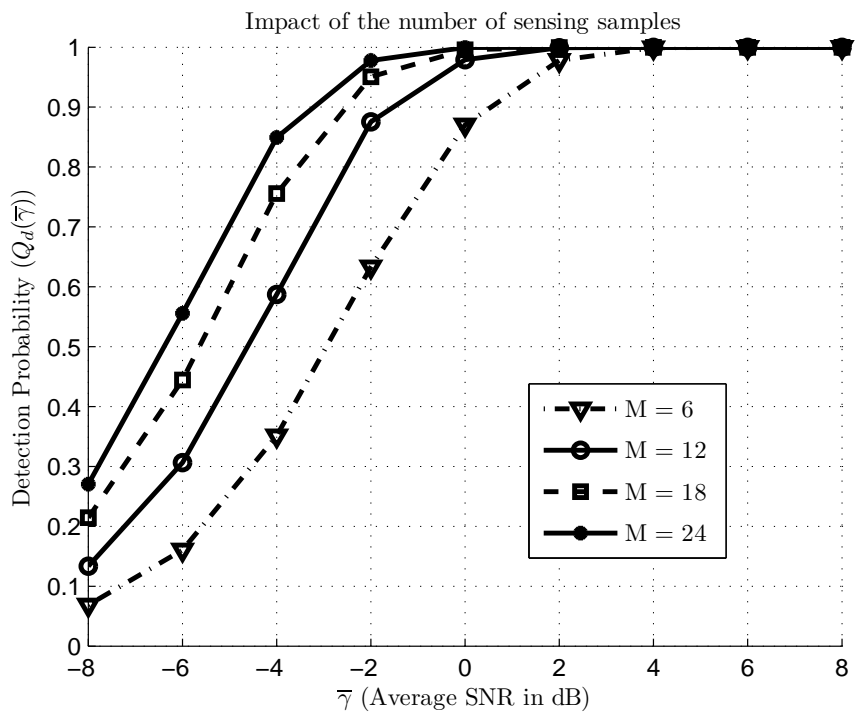}
\caption{SNR gain obtained by increasing the number of sensing samples.}
\label{fig_sim}
\end{figure}

\begin{figure}[!h]
\centering
\includegraphics[width=5in]{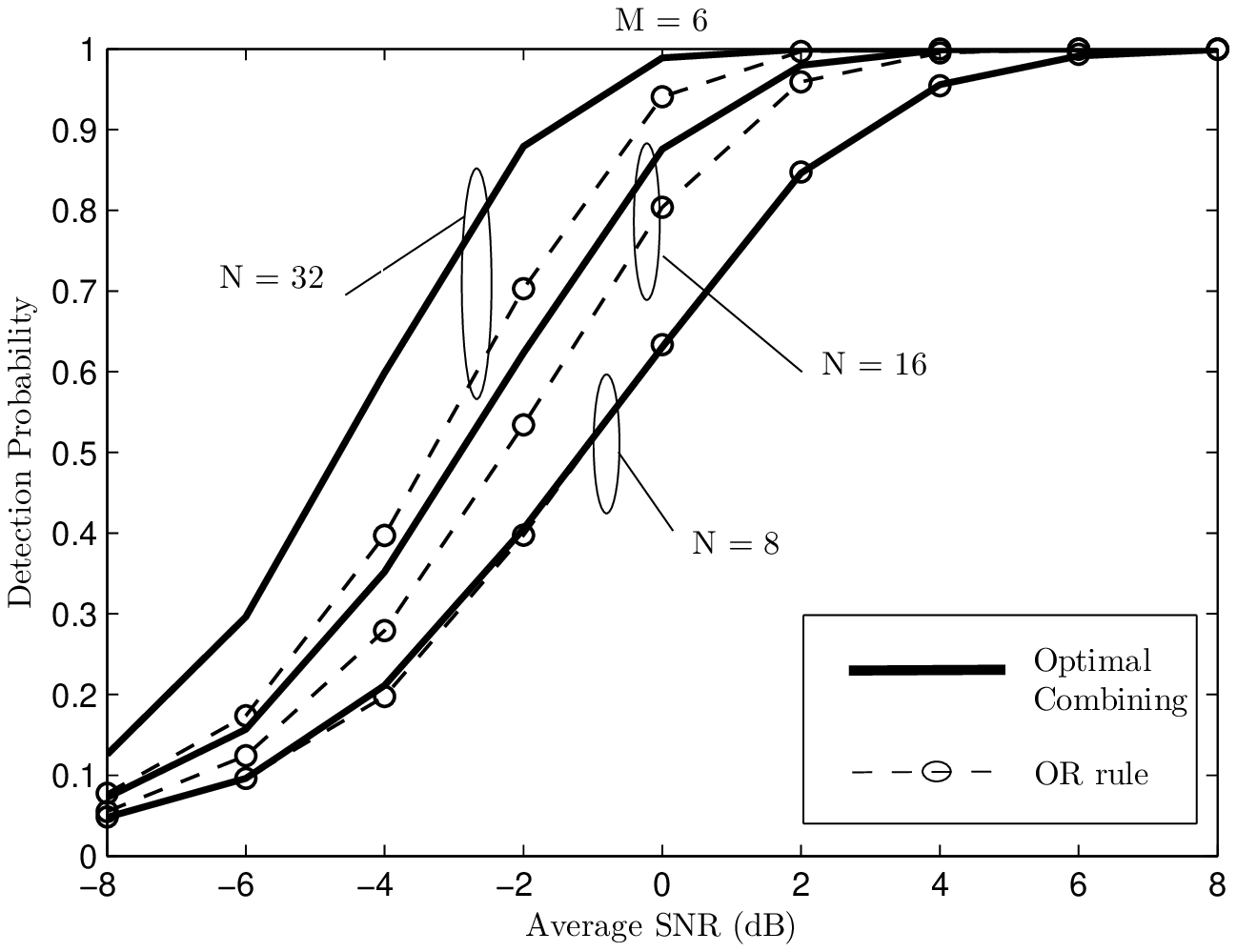}
\caption{Optimality of the proposed fusion rule in networks with large number of cooperating users.}
\label{fig_sim}
\end{figure}

\section{Conclusion}
In this paper, we proposed a globally optimal hard decisions fusion scheme for cooperative spectrum sensing. This problem has been always known for being complex and mathematically intractable. We have proved that the optimal local and global Neyman-Pearson thresholds can be obtained by a simple convex optimization problem. This is achieved by  utilizing the mathematical representation of the global detection and false alarm probabilities in terms of a regularized incomplete beta function. The log-concavity of global detection probability as a function of the global threshold paves the way for constructing a convex objective function. The proposed algorithm has a complexity of $\mathcal{O}(\log_{2}(N))$. Simulation results verify the optimality of the proposed scheme. It is shown that the globally optimal scheme offers significant gain only when the number of cooperating users is large. Otherwise, one can use a simple OR-rule.



\begin{thebibliography}{}
%
%
\bibitem{IEEEhowto:kopka}
S.~Haykin, "Cognitive radio: brain-empowered wireless communications," \emph{IEEE Journal on Selected Areas in Communications}, vol. 23, pp. 201-220, Feb. 2005.
\bibitem{IEEEhowto:kopka2}
A.~Ghasemi and E.~Sousa, "Collaborative spectrum sensing for opportunistic access in fading environments," \emph{First IEEE International Symposium on New Frontiers in Dynamic Spectrum Access Networks}, pp. 131 ֠136, Nov. 2005.
\bibitem{IEEEhowto:kopka3}
Wei Zhang, R.~K.~Mallik and K.~Ben Letaief, "Cooperative Spectrum Sensing Optimization in Cognitive Radio Networks," \emph{Proceedings of IEEE International Conference on Communications (ICCҰ8)}, pp. 411 ֠3415, May 2008.
\bibitem{IEEEhowto:kopka11}
Jun Ma, Guodong Zhao  and  Ye Li, "Soft Combination and Detection for Cooperative Spectrum Sensing in Cognitive Radio Networks," \emph{IEEE Transactions on Wireless Communications}, vol. 7, pp. 4502-4507, Nov. 2008.
\bibitem{IEEEhowto:kopka4}
J.~N.~Tsitsiklist, "Decentralized Detection by a Large Number of Sensors," \emph{Mathematics of Control, Signals and
Systems}, vol. 1, no. 2, pp. 167ֱ82, 1988. 
\bibitem{IEEEhowto:kopka5}
Zhi Quan, Shuguang Cui, H.~Vincent Poor, and Ali H. Sayed, "Collaborative Wideband Sensing for Cognitive Radios," \emph{IEEE Signal Processing Magazine}, vol. 25, no. 6, pp. 60-73, Nov. 2008. 
\bibitem{IEEEhowto:kopka5}
Imad Y. Hoballah and Kumar Varshney, "Neyman-Pearson detection wirh distributed sensors," \emph{ 1986 25th IEEE Conference on Decision and Control,} Syracuse University, Syracuse, New York, Dec. 1986, pp. 237-241.    
\bibitem{IEEEhowto:kopka6}
Dongliang Duan, Liuqing Yang and Louis L.~Scharf, "Optimal Local Detection for Sensor Fusion by Large Deviatiob Analysis," \emph{20th European Signal Processing Conference (EUSIPCO 2012)}, Bucharest, Romania, August 27 - 31, 2012, pp. 744-748. 
\bibitem{IEEEhowto:kopka7}
Junyang Shen, Tao Jiang, Siyang Liu, and Zhongshan Zhang, "Maximum Channel Throughput via Cooperative Spectrum Sensing in Cognitive Radio Networks," \emph{IEEE Transactions on Wireless Communications}, vol. 8, no. 10, pp. 5166-5175, Oct. 2009.
\bibitem{IEEEhowto:kopka8}
Yunfei Chen, "Analytical Performance of Collaborative Spectrum Sensing Using Censored Energy Detection," \emph{IEEE Transactions on Wireless Communications}, vol. 9, no. 12, pp. 3856-3865, Dec. 2010.
\bibitem{IEEEhowto:kopka12}
J.~Shen, S.~Liu, L.~Zeng, G.~Xie, J.~Gao and Y.~Liu, "Optimisation of cooperative spectrum sensing in cognitive radio network,"  \emph{IET Communications}, vol. 3, pp. 1170-1178, March 2008.
\bibitem{IEEEhowto:kopka10}
S.~Chaudhari, J.~Lunden, V.~Koivunen, H.~V.~Poor, "Cooperative Sensing With Imperfect Reporting Channels: Hard Decisions or Soft Decisions?," \emph{IEEE transactions on Signal Processing}, vol. 60, pp. 18-28, Jan. 2012.      
\bibitem{IEEEhowto:kopka14}
Milton Abramowitz and Irene Stegun, "Handbook of Mathematical Functions with Formulas, Graphs, and Mathematical Tables," \emph{Dover Publications}, ISBN 0-486-61272-4, 1964. 
\bibitem{IEEEhowto:kopka15}
Yin Sun, A. Baricz, and Shidong Zhou, "On the monotonicity, log-concavity and tight bounds of the generalized Marcum and Nuttall Q−functions" \emph{IEEE Transactions on Information Theory}, vol. 56, no. 3, pp. 1166 - 1186, March 2010. 
\bibitem{IEEEhowto:kopka15}
D.Karp and S.M. Sitnik, "Log-convexity and log-concavity of hypergeometric-like functions," \emph{Journal of Mathematical Analysis and Applications}, vol. 364, issue 2,  pp. 384–394, Apr. 2010.
\bibitem{IEEEhowto:kopka15}
Arpad Baricz, Saminathan Ponnusamy, and Matti Vuoeinen, "Functional Inequalities for Modified Bessel Functions," \emph{Journal of Mathematical Analysis and Applications}, vol. 364, issue 2,  pp. 384–394, Mar. 2011.
\bibitem{IEEEhowto:kopka15}
Mark Bagnoli and Ted Bergstrom, "Log-Concave Probability and Its Applications," \emph{Economic Theory, Springer}, vol. 26, no. 2, pp. 445-469, Aug. 2005.
\bibitem{IEEEhowto:kopka15}
Stephen Boyd and Lieven Vandenberghe, "Convex Optimization," \emph{Cambridge University Press}, 2004.
\end{thebibliography}


\end{document}